\documentstyle[12pt,prepriop]{article}
\begin{document}
\begin{center}
\bf   THE DYNAMICS OF VORTEX STRUCTURES AND STATES OF CURRENT IN 
    PLASMA-LIKE FLUIDS AND THE ELECTRICAL EXPLOSION OF CONDUCTORS:\\
1. The model of a non-equilibrium phase transition. 
\end{center}
\begin{center}
N B Volkov and A M Iskoldsky\\
Russian Academy of Science, Ural Division\\
Institute of Electrophysics\\
34 Komsomolskaya St.\\
Yekaterinburg 620219,\\ 
Russia
\end{center}
\vspace{3.5cm}
\begin{center}
\bf The Dynamics of Vortex Structures and States of Current: 1
\end{center}
\newpage
\begin{center}
{\bf   THE DYNAMICS OF VORTEX STRUCTURES AND STATES OF CURRENT IN 
    PLASMA-LIKE FLUIDS AND THE ELECTRICAL EXPLOSION OF CONDUCTORS:\\    
          1. The model of a non-equilibrium phase transition}\\
\bigskip 
                     N B Volkov and A M Iskoldsky
\end{center}
\begin{center}
\bf Abstract
\end{center}
A set of equations according to which the conducting  medium  consists 
of two fluids - laminar and vortex, has been obtained in  the  present 
paper by transforming MHD equations. In a similar way,  an  electronic 
fluid is assumed to consist of a laminar  and  a  vortex  fluid.  This 
system  allows  one  to  study  the  formation  and  the  dynamics  of 
large-scale hydrodynamic fluctuations. From this model a  model  of  a 
non-equilibrium phase  transition  belonging  to  a   class   of   the 
Lorenz-type models has been developed  [Lorenz  E  N  1963  J.  Atmos. 
Sci. {\bf 20} 130]. Vortex structures resulting in the increase in  an 
effective resistance of the conducting  medium  and  the  interruption 
of current have been shown to appear even at constant transport 
coefficients in a laminar electronic fluid. Critical exponents of the 
parameters of an order (amplitudes), which for a direct current 
coincide with the critical exponents in the Lorenz model, have been 
found. A spatial scale of the structure described by the theory is  in 
good  agreement  with  experiment.  A  further  evolution  of   vortex 
structures has been shown to occur by splitting the spatial  scale.  A 
similarity, according to which the  following  sequence  of  splitting 
takes place: {\({k_0}{\rightarrow}{k_1}=0.5{k_0}{\rightarrow}
{k_2}=2{k_0}{\rightarrow}{k_3}=2{k_2}\)}, etc. has been 
hypothesized.
\newpage
\section{Introduction}
\par
The present series of three  papers  describes  the  dynamics  of 
vortex structures and states of current in  plasma-like  fluids  as  a 
magnetohydrodynamic  (MHD)  approximation.   Vortex   structures   are 
large-scale (hydrodynamic) fluctuations, obeying the condition
{\(kL\geq1\)}, where {\(k\)} and {\(L\)} are, respectively, the
wave number and the characteristic size (for instance, the sample size). 
Criteria for the classification of fluctuations into small-scale
(kinetic) with {\(kL\ll1\)} and 
large-scale are discussed in [1]. In
addition, a local thermodynamic 
equilibrium (LTE) is valid, and local kinetic transport coefficients 
({\(\sigma\)} is the electric conductivity, {\(\kappa\)} is the 
thermal conductivity, {\(\eta\)} is 
the shear viscosity) are still determined by kinetic
fluctuations. 
\par
Below in section 3 and also in the second paper of the series  we 
show that even in the case of a  constant  transport  coefficient,  an 
emergence of vortex hydrodynamic structures results in  a  spontaneous 
breaking of a symmetry of a laminar, in the initial state,  "electron" 
fluid and the appearance of vortex current structures. As a
consequence, the current in a cylindrical conductor is interrupted, i.e. an 
effective resistance of the conductor
{\({R_{eff}}{\rightarrow}{\infty}\)} and a region of 
negative differential resistance appears on the $UI$-characteristic. This 
process is singular in time and for its realization  it  is  necessary 
that the external electric circuit should  be  supercritical  to  some 
extent, which is a characteristic property of a non-equilibrium  phase 
transition (NPT) [2].
\par
Section 3 presents a simple NPT model for a cylindrical
conductor. To develop it we used set of magnetohydrodynamic equations 
(MHD) obtained in section 2. We also  showed  that  the  formation of 
vortex structures in the medium, from the view point  of  an  external 
observer results in an effective alteration in  kinetic  coefficients, 
and local kinetic coefficients determined by  small-scale  fluctuation 
as before. 
\par
In the second paper of the given series using computer simulation 
we studied the dynamics of vortex structures  and  current  states  in 
electrophysical systems where the NPT model from section 3 is used  as 
a model of a non-linear element (NE). 
\par
In the 3rd paper we discuss and give a qualitative explanation of 
the electrical explosion in conductors (EEC) [3], 
[4], which is an example of 
a phenomenon where vortex structures play, in our opinion, the 
dominant role (see [5], which draws attention 
to the  analogy  between 
initial stages of EEC and the turbulence in an incompressible liquid). 
During an electrical explosion the conductor firstly breaks down  into 
transverse strata (axial structurization) and then it expands  forming 
a low temperature plasma with a condensed disperse phase (CDP) by  the 
end of the explosion. It has been established experimentally 
[4] that 
the conductor stratification takes a time that is less than a
characteristic hydrodynamic time, particularly it is less  than  the  sound 
time {\(t_s={r_0}{c_s}^{-1}\)} ({\(r_0\)}, {\(c_s\)} 
are the conductor radius and the sound 
velocity respectively). The above fact  and  other  experimental  evidence 
earlier considered  anomalous,  find  a  natural  explanation  in  the 
framework of our model. 
\section{Mathematical model}
\par
For the initial mathematical model we use hydrodynamic equations of  a 
conducting liquid and Maxwell equations [6], 
restricting ourselves  to 
the MHD approximation. We assume that there is LTE in the 
system. As 
noted in [7], physical processes resulting in
the infringement of LTE make 
a major grade contribution to the bulk viscosity; therefore proceeding 
from the assumption of LTE, the bulk viscosity is  not  considered  in 
the set of equations below (the existence of large-scale vortex 
excitations in the medium leads to physical effects that can be  explained 
as the influence of an effective bulk viscosity). 
\par
In this case the set of MHD equations becomes 
\begin{equation}
\frac{\partial\rho}{\partial t} + ({\bi{u}}\cdot{\nabla}){
\rho} + \rho({\nabla}\cdot{\bi{u}}) = 0,
\end{equation}
\begin{eqnarray}
\lefteqn{\rho\Biggl(\frac{\partial{\bi{u}}}{\partial t} -  
[\bi{u},[{\nabla},{\bi{u}}]] + {\nabla}
{\frac{u^2}{2}}\Biggr)=-{\nabla}P +} 
                    \nonumber\\ 
& & +\frac{1}{4\pi}[[{\nabla},{\bi{H}}],{\bi{H}}] + 
{\nabla}\cdot{\hat {\bss{S}}},
\end{eqnarray}
\begin{eqnarray}
\rho{c_\rho}\Biggl(\frac{\partial T}{\partial t} + 
({\bi{u}}\cdot{\nabla})T\Biggr) & = & -P_T({\nabla}\cdot
{\bi{u}}) + \frac{\nu_m}{4\pi}([{\nabla},{\bi{H}}])^2 + \nonumber\\
& & +Sp({\hat {\bss{S}}}\cdot({\nabla}\otimes{\bi{u}})^T) + 
({\nabla}\cdot\kappa{\nabla}T),
\end{eqnarray}
\begin{equation}
[{\nabla},{\bi{E}}] = -\frac{1}{c}\frac{\partial
{\bi{H}}}{\partial t},
\end{equation}
\begin{equation}
[{\nabla},{\bi{H}}] = \frac{4\pi}{c}{\bi{j}},
\end{equation}
\begin{equation}
({\nabla}\cdot{\bi{H}}) = 0,
\end{equation}
\begin{equation}
({\nabla}\cdot{\bi{E}}) = 4{\pi}{\delta}{\rho}_e,
\end{equation}
where {\(\rho\)}, {\(\bi{u}\)}, {\(T\)}, {\(P\)}, 
{\(\bi{E}\)}, {\(\bi{H}\)}, {\(\delta{\rho_e}\)},
{\(\bi{j}\)} are, respectively  the  local  density, 
the hydrodynamic velocity, the temperature, the pressure, the electric 
intensity, the magnetic intensity, the density of a nonbalanced 
electric charge (in the MHD approximation 
{\({\langle}{\rho}_e{\rangle} = 0\)}; Eq. (7) is used below 
to find the deviation $\delta{\rho_e}$  
from the mean charge  value  induced  by  an 
effective polarization of  the  conducting  medium)  and  the  current 
density. To find the latter, we use the generalized Ohm's law  in  the 
simplest form 
\begin{equation}
{\bi{j}} = \sigma({\bi{E}} + {\frac{1}{c}}[{\bi{u}},{\bi{H}}]);
\end{equation}
{\(P_T = T(\partial P/\partial \rho)_T\)}; {\({\hat {\bss{S}}} =
2\eta({\hat {\bss{U}}} - 3^{-1}({\nabla}\cdot{\bi 
{u}}){\hat {\bss{I}}})\)} 
is the  tensor  of  a  viscous 
stress, ${\hat {\bss{U}}} = 0.5(({\nabla}\otimes{\bi{u}}) + 
({\nabla}\otimes{\bi{u}})^T)$, ${\hat {\bss{I}}}$ is the unit tensor 
(the symbol $"\sf T"$ 
denotes a transposition; $"\cdot"$ is the internal (scalar) 
vector product, $"\otimes"$ is the tensor product of vectors; 
$[{\bi{a}},{\bi{b}}]$ is the vector product of ${\bi{a}}$ 
and ${\bi{b}}$ vectors;${\nabla} = {\nabla_i}{\bi{e}^i}$ is the 
gradient operator); $c_{\rho}$ is the heat 
capacity per unit mass; ${\nu}_m = c^2(4\pi\sigma)^{-1}$ is 
the  magnetic  viscosity, $c$ is 
the velocity of light.
\par
To close the set of Eqs. (1) - (8), it is necessary to 
specify  initial 
and boundary conditions. In our case, it  is  sufficient  to  set  the 
value of a normal 
\begin{equation}
P - {\hat {\bss{S}}}_{nn} = 0
\end{equation}       
and a tangential
\begin{equation}                                                      
{\hat {\bss{S}}}_{n\tau} = 0
\end{equation}       
component of the stress tensor and the value of the magnetic intensity 
on a free boundary between medium and vacuum ($\hat
{\bi{S}}_n={\hat {\bss{S}}}\cdot{\bi{n}}$, $\bi{n}$ is the 
normal vector to the conductor surface). 
\par
We rewrite Eqs.(1) - (8) to the form where large-scale vortex 
excitations can be explicitely represented. It is remarkable 
that a significant role in 
their formation is played by a surface that restricts the volume
occupied by a continuous conducting medium. This surface enables 
distinguishing vector fields (of  hydrodynamic  velocity  and  current 
density) which are connected to themselves inside and 
outside this  volume  or 
end on the boundary  (conventionally  called  "laminar"  and  "vortex" 
vector fields). 
\par
Actually, let
\begin{equation}
{\bi{u}} = {\bi{u}}_{\rho} + {\bi{u}}_{\omega}, 
\end{equation}
with
\begin{equation}                                                                   
({\nabla}\cdot{\bi{u}}_{\omega}) = 0,\qquad
[{\nabla},{\bi{u}}_{\rho}] = 0,
\end{equation}
where ${\bi{u}}_{\rho}$, ${\bi{u}}_{\omega}$ are the laminar and the  
vortex  vector  fields  of 
hydrodynamic velocity respectively. We write the motion equation of  a 
vortex liquid in the following form:
\begin{eqnarray}      
\lefteqn{\rho\Biggl(\frac{\partial {\bi{u}}_{\omega}}{\partial t} - 
[{{\bi{u}}_{\omega}},[{\nabla},{\bi{u}}_{\omega}]] + 
{\nabla}{\frac{{u_{\omega}}^2}{2}}\Biggr) = }
                           \nonumber\\  
& & =-{\nabla}{(P_{\omega} - P_m)} + 
{\frac{1}{4\pi}[[{\nabla}, \bi{H}], \bi{H}]} + 
{\nabla}\cdot{{\hat {\bss{S}}}_{\omega}}, 
\end{eqnarray}
where $P_m = {H^2}(8\pi)^{-1}$ is the magnetic pressure; 
${{\hat {\bss{S}}}_{\omega}}=2{\eta}{{\hat {\bss{U}}}_{\omega}}$; 
${{\hat {\bss{U}}}_{\omega}}=0.5({\nabla}\otimes
{{\bi{u}}_{\omega}} + ({\nabla}\otimes{{\bi{u}}_{\omega}})^{\sf
T})$; 
$P_{\omega}$ is the  hydrodynamic pressure the spatial 
distribution of which is derived from the 
solution of the Poisson-type equation found by using 
the first relation of Eq.(12) in Eq.(13): 
\begin{eqnarray}
\lefteqn{\Delta(P_{\omega} - P_m) - {\nabla}(\ln{\rho})\cdot
{\nabla}(P_{\omega} - P_m) = \rho{\nabla}\cdot\biggl(
- {\nabla}{\frac{{u_{\omega}}^2}{2}} + }
                      \nonumber\\
& & + [{\bi{u}}_{\omega},[{\nabla},{\bi{u}}_{\omega}]] + 
{{\rho}^{-1}}\bigl((4\pi)^{-1}[{\nabla},[{\nabla},\bi{H}]]
+ {\nabla}\cdot{{\hat {\bss{S}}}_{\omega}}\bigr)\biggr).
\end{eqnarray}
\par
By substituting (12) into (1) - (2) with the consideration of 
Eq.(13), we obtain: 
\begin{equation} 
\frac{\partial \rho}{\partial t} + ({{\bi{u}}_{\rho}}\cdot
{\nabla})\rho + \rho({\nabla}\cdot{{\bi{u}}_{\rho}}) = 
-({{\bi{u}}_{\omega}}\cdot{\nabla})\rho, 
\end{equation}
\begin{eqnarray}
\lefteqn{\rho\Biggl(\frac{\partial {\bi{u}}_{\rho}}{\partial t} 
+ {\nabla}{\frac{{u_{\rho}}^2}{2}}\Biggr) = 
-{\nabla}(P_{\rho} + P_m ) - }
                   \nonumber\\
& & -\rho({\nabla}({{\bi{u}}_{\rho}}\cdot{{\bi{u}}_{\omega}}) - 
[{\bi{u}}_{\rho},[{\nabla},{\bi{u}}_{\omega}]]) + 
{\nabla}\cdot{{\hat {\bss{S}}}_{\rho}},
\end{eqnarray}
where $P_{\rho} = P - P_{\omega}$; ${{\hat
{\bss{S}}}_{\rho}}=2{\eta}({{\hat {\bss{U}}}_{\rho}}-{3^{-1}}
({\nabla}\cdot{{\bi{u}}_{\rho}}){\hat {\bss{I}}})$; 
${{\hat {\bss{U}}}_{\rho}}=0.5({\nabla}\otimes
{{\bi{u}}_{\rho}} + ({\nabla}\otimes
{{\bi{u}}_{\rho}})^{\sf T})$. From the viewpoint of 
an external observer the second term in the right side of 
Eq. (16) has a physical sense of an effective volume viscosity. 
\par
As consistent with the second relation from (12), Eq. (16) 
has a form: 
\begin{equation}
\frac{\partial {\bi{u}}_{\rho}}{\partial t} = - {\nabla}
(\frac{u^2}{2} + \chi),
\end{equation}
where $\chi$ is the effective chemical potential the spatial 
distribution of which is derived from the Poisson-type equation 
found by using expressions (12) and (14) in Eq. (16):
\begin{eqnarray}
\lefteqn{{\Delta \chi} = {\nabla}\cdot\Biggl({\rho}^{-1}
\biggl({\nabla}(P-{\frac{4}{3}}\eta({\nabla}\cdot
{{\bi{u}}_{\rho}})) - (4\pi)^{-1}[[{\nabla},\bi{H}],\bi{H}] 
+ }
                 \nonumber\\
& & +[{\nabla},\eta[{\nabla},{\bi{u}}_{\omega}]] 
+ 2({\nabla}\cdot{{\bi{u}}_{\rho}}){\nabla}\eta - 
2{\nabla}\cdot({\nabla}\otimes{\bi{u}})^{\sf T}\biggr) - 
[\bi{u},[{\nabla},{\bi{u}}_{\omega}]]\Biggr).
\end{eqnarray}
In the region of developed large-scale fluctuations 
the constraint equation (18) should be used as an 
effective equation of the  state  of 
liquid. For a homogeneous incompressible liquid with $\eta=0$, 
$|{\bi{H}}|=0$ and $|{{\bi{u}}_{\omega}}|=0$, 
the solution Eq. (18) has 
a form: $\chi=P{\rho}^{-1}$.
\par
The boundary conditions are rewritten as 
\begin{eqnarray}
P_{\rho} - {\hat {\bss{S}}}_{\rho,nn} & = & 
-(P_{\omega} - {\hat {\bss{S}}}_{\omega,nn}),   \\
{\hat {\bss{S}}}_{\rho,n\tau} & = & 
-{\hat {\bss{S}}}_{\omega,n\tau}. 
\end{eqnarray}
Hence, the existence of large-scale vortex excitations in  the  liquid 
results in the appearance of back pressure and tangent stress  on  the 
free boundary with a vacuum, which prevents from a change  in  the 
shape of the volume occupied by the liquid. 
\begin{sloppypar}
Vortex current excitations can also appear in plasma-like media, 
therefore we transform Maxwell equations (4) - (7), taking Ohm's law 
into account, in the form of (8). Let 
${\bi{j}} = {\bi{j}}_{\rho} + {\bi{j}}_{\omega}$, with 
\end{sloppypar}
\begin{equation}
\int_{\Sigma}{\bi{j}}_{\rho}\cdot{\bi{ds}} = I,\qquad
\int_{\Sigma}{\bi{j}}_{\omega}\cdot{\bi{ds}} = 0
\end{equation}
($I$ is the total current, $\Sigma$ is  the  surface  
restricting  the  volume 
occupied by the conducting medium). Then 
\begin{eqnarray}
{\bi{j}}_{\rho} & = & \sigma({\bi{E}}_{\rho} + 
c^{-1}[{\bi{u}},{\bi{H}}_{\rho}]),      \\
{\bi{j}}_{\omega} & = & \sigma({\bi{E}}_{\omega} + 
c^{-1}[{\bi{u}},{\bi{H}}_{\omega}]),
\end{eqnarray}
\begin{equation}
{[{\nabla},{\bi{E}}_{\rho}]} =  
-c^{-1}\frac{\partial {\bi{H}}_{\rho}}{\partial t},
\end{equation}
\begin{equation}
{[{\nabla},{\bi{E}}_{\omega}]} =  
-c^{-1}\frac{\partial {\bi{H}}_{\omega}}
{\partial t},
\end{equation}
\begin{eqnarray}
{[{\nabla},{\bi{H}}_{\rho}]} & = & 
4{\pi}c^{-1}{\bi{j}}_{\rho},        \\
{[{\nabla},{\bi{H}}_{\omega}]} & = & 
4{\pi}c^{-1}{\bi{j}}_{\omega},
\end{eqnarray}
\begin{eqnarray}
({\nabla}\cdot{\bi{E}}_{\rho}) + 
({\nabla}\cdot{\bi{E}}_{\omega}) & = & 
4{\pi}{\delta}{\rho}_e ,    \\ 
({\nabla}\cdot{\bi{H}}_{\rho}) + 
({\nabla}\cdot{\bi{H}}_{\omega}) & = & 0.  
\end{eqnarray} 
\par
In the absence of vortex current excitations, i.e. at 
$|{\bi{j}}_{\omega}| = 0$,
$({\nabla}\cdot{\bi{H}}_{\rho} ) = 0$. Therefore, 
to exclude nonphysical  solutions,  we  require 
that
\begin{eqnarray} 
{({\nabla}\cdot{\bi{H}}_{\rho})} = 0,     \\ 
{({\nabla}\cdot{\bi{H}}_{\omega})} = 0.
\end{eqnarray} 
\par
Then a set of diffusion equations ${\bi{H}}_{\rho}$ and 
${\bi{H}}_{\omega}$ follows from Eqs. (21) 
- (27) and Eqs. (30) - (31):
\begin{eqnarray} 
\lefteqn{\frac{\partial {{\bi{H}}_{\rho}}}{\partial t} + 
({{\bi{u}}_{\rho}}\cdot{\nabla}){\bi{H}}_{\rho} = 
({\bi{H}}_{\rho}\cdot{\nabla}){\bi{u}}_{\rho} - 
{\bi{H}}_{\rho}({\nabla}\cdot{\bi{u}}_{\rho}) + }
         \nonumber\\
& & +\{({\bi{H}}_{\rho}\cdot{\nabla}){\bi{u}}_{\omega} - 
({\bi{u}}_{\omega}\cdot{\nabla}){\bi{H}}_{\rho}\} + 
\nu_m{\Delta}{\bi{H}}_{\rho} - [{\nabla}{\nu_m},
[{\nabla},{\bi{H}}_{\rho}]],
\end{eqnarray}
\begin{eqnarray}
\lefteqn{\frac{\partial {{\bi{H}}_{\omega}}}
{\partial t} + 
({{\bi{u}}_{\omega}}\cdot{\nabla})
{\bi{H}}_{\omega} = 
({\bi{H}}_{\omega}\cdot{\nabla})
{\bi{u}}_{\omega} - 
{\bi{H}}_{\omega}({\nabla}\cdot{\bi{u}}_{\rho}) + }
         \nonumber\\
& & +\{({\bi{H}}_{\omega}\cdot{\nabla}){\bi{u}}_{\rho} - 
({\bi{u}}_{\rho}\cdot{\nabla}){\bi{H}}_{\omega}\} + 
\nu_m{\Delta}{\bi{H}}_{\omega} - [{\nabla}{\nu_m},
[{\nabla},{\bi{H}}_{\omega}]].
\end{eqnarray}
\par
The terms between the  brackets  in  (32)  and  (33)  consider  the 
influence of laminar and vortex vector fields of the hydrodynamic velocity 
on the alteration in magnetic fields. This influence can  be  
recognized 
by an external observer as an  effective  variation  in  the  magnetic 
diffusion factor. It should be also noted that a direct measurement of 
the field ${\bi{H}}_{\omega}$ is impossible due to condition
(21). However, it contributes to an effective voltage drop 
across a conductor, thus determining the total current $I$ (see section 3). 
\par
The energy balance equation (3) remains unaltered, the  work  of 
pressure forces to expand (compress) the matter  being  determined  by 
the field ${\bi{u}}_{\rho}$, and local kinetic transport factors 
by - small-scale fluctuations. 
\begin{sloppypar}
Let us make some comments. The equations obtained are a 
phenomenological set of equations for a two-liquid hydrodynamic 
description of a "heavy" (conducting fluid) and a "light" fluid of 
(magnetic field). In this framework a simple model of  
a non-equilibrium  phase  transition 
(section 3) has been developed and the dynamics of vortex  structures 
and states of current in electrophysical circuits  has  been  studied. 
While deriving Eqs.(13) - (18) and (32) - (33) we did not use additional 
physical considerations which were beyond  the  applicability  of  the 
initial  model  (1) - (8).  In  addition,  an  essential  
condition  of  the consistency of motions determined 
by this system  is  to  satisfy  the 
condition of constraint (18), which is  similar  to  the  condition  of 
deformation compatibility in the elastic theory. From the view point of 
an external observer, characteristics of the field 
${\bi{u}}_{\omega}$ can be  regarded 
as additional thermodynamic  variables  which  enter  a  thermodynamic 
potential as independent variables. To describe the  dynamics  of  the 
latter, it is necessary to derive relaxation-type  equations 
[7], [8] 
and to use experimental data for finding a  relaxation  time.  In  the 
framework of the field approach developed in  this  paper,  the  above 
problem no longer arises since relaxation times will appear only  when 
Eqs. (13) - (14) are reduced to relaxation-type equations.
\end{sloppypar} 
\section{The model of a non-equilibrium phase transition}
\par
A homogeneous incompressible liquid is used  in  the  development 
of the NPT model. This fluid occupies the region of a fixed  size  (as 
is shown in [5], it is correct for the 
initial stage of EEC; discussed 
below is a liquid conducting cylinder with the radius $r_0$  and  the 
length $l\gg r_0$). In this case $|{\bi{u}}_{\rho}| = 0$ 
and the  compatibility  condition 
(17) is satisfied identically. Moreover, the same as in 
[5], we  assume 
local kinetic coefficients to be constant and consequently, 
we do  not  take 
into consideration, in a first approximation,  the  influence  of  the 
conductor heating on the dynamics of vortex structures and  states  of 
current. 
\par
Then the set of Eqs.(11) - (16) and (32) -(33) is  written  in  the 
following form (below we do not use the index $"\omega"$ since 
we  discuss  a 
vortex fluid; moreover, instead of Eq. (32) and Eq. (33),  we  use  a 
diffusion equation for the total magnetic field ${\bi{H}}$):
\begin{equation}  
({\nabla}\cdot{\bi{u}}) = 0 ; 
\end{equation}
\begin{equation}
\frac{\partial {\bi{u}}}{\partial t} + 
({\bi{u}}\cdot{\nabla}){\bi{u}} = 
-{\rho_0}^{-1}{\nabla}P + (4{\pi}{\rho}_0)^{-1}
[[{\nabla},{\bi{H}}],{\bi{H}}] + \nu{\Delta}{\bi{u}}; 
\end{equation}
\begin{equation}
\frac{\partial {\bi{H}}}{\partial t} + 
({\bi{u}}\cdot{\nabla}){\bi{H}} = ({\bi{H}}\cdot
{\nabla}){\bi{u}} + {{\nu}_m}{\Delta}{\bi{H}}, 
\end{equation}
where $\nu = \eta{\rho_0}^{-1}$  is the kinematic viscosity 
and $\Delta$ is the Laplacian. As 
follows from Eq. (34), ${\bi{u}} = [{\nabla},{\bi{A}}]$ 
(${\bi{A}}$ is the vector  potential  of  the 
velocity, which is known to be  accurate  within  the  gradient  of  a 
scalar function; below we use  Coulomb  gauge of  the  vector 
potential: $({\nabla}\cdot{\bi{A}}) = 0$).
\par
We direct the $z$ axis along the axis of the conductor and we  make 
use of the azimuthal symmetry, setting 
${\bi{u}} = \{{u_r}(r,z,t), 0, {u_z}(r,z,t)\}$, 
${\bi{H}} = \{0, H(r,z,t), 0\}$, ${\bi{A}} = \{0, {\psi}(r,z,t),
0\}$. From Eq.(34)  we  then 
find $u_r = -{\partial \psi}/{\partial z}$, 
$u_z = {\partial (r\psi)}/{r{\partial r}}$. We look 
for a solution of Eq.(36) 
in the form $H(r,z,t) = {H_1}(r,t) + h(r,z,t)$, where 
${H_1}(r,t) = 2i(t)r(c{r_0}^2)^{-1}$, 
i.e. while finding $H_1$ , the current density is assumed to be
homogeneous in the cross-section of the conductor. In this 
case the boundary 
conditions for Eq.(36) are satisfied automatically, since 
${H_1}(r_0,t) = 2i(t)(cr_0)^{-1}$. Correspondingly, the 
boundary conditions for the field 
$h(r,z,t)$ at $r = 0$ and $r = r_0$  are  zero: $h(0,z,t) = 
h(r_0,z,t) =  0$. 
Applying the vector operation $[{\nabla},\quad]$ to Eq.(35) 
and making use of  the 
above assumptions, we  derive  a  set  of  differential  equations  in 
partial derivatives for $\psi$ and $h$:
\begin{eqnarray}
\lefteqn{\frac{\partial (\Delta{\psi} - {\psi}r^{-2})}
{\partial t} = - \frac{\partial (\psi,\Delta{\psi} -
{\psi}r^{-2})}{\partial (r,z)} + {\frac{2(\Delta{\psi} - 
{\psi}r^{-2})}{r}}{\frac{\partial \psi}{\partial z}} + }
           \nonumber\\
& & +R{\frac{{\nu}{\nu_m}}{H_0{r_0}^3}}{\frac{i}{I_0}}
{\frac{\partial h}{\partial z}} + \nu[\Delta(\Delta{\psi} - 
{\psi}r^{-2}) - r^{-2}(\Delta{\psi} - {\psi}r^{-2})],
\end{eqnarray}
\begin{equation}
\frac{\partial h}{\partial t} = -\frac{\partial (\psi,h)}
{\partial (r,z)} + {\frac{H_0}{r_0}}{\frac{i}{I_0}}
{\frac{\partial \psi}{\partial z}} + {\nu_m}({\Delta}h - 
\frac{h}{r^2}),
\end{equation}
where $$\frac{\partial (a,b)}{\partial (r,z)} = 
{\frac{\partial (ra)}{r{\partial r}}}
{\frac{\partial b}{\partial z}} - {\frac{\partial a}{\partial
z}}{\frac{\partial (rb)}{r{\partial r}}},$$ 
$R = {H_0}^2{r_0}^2(2{\pi}{\rho_0}{\nu_m}{\nu})^{-1} = 
{v_A}^2{r_0}^2({\nu_m}{\nu})^{-1} = {Pe_m}^2{s^{-1}}$ is 
the Rayleigh number, $v_A = {H_0}(2{\pi}{\rho_0})^{-{1/2}}$ is 
the Alfven velocity, $Pe_m = {v_A}{r_0}{\nu_m}^{-1}$ is 
the  magnetic  Peclet  number, 
$s = {\nu}{\nu_m}^{-1}$, $H_0 = 2{I_0}(cr_0)^{-1}$, $I_0$ 
is the  characteristic  current  value 
dependent on the type of energy source. At the direct current  
$i(t) = I_0 = const$, the set of Eqs.(37) - (38), as shown 
in [5], agrees accurately to 
the replacement of a Cartesian coordinate system by a cylindrical  one 
and the replacement of the heat conduction equation for the  diffusion 
of a magnetic field, with the set of Saltzman equations 
[9] in the theory of the Benard effect 
[10]. 
\par
A second term in the right-hand  side  of  Eq.(37)  leads to the 
breakdown of azimuthal symmetry (leads to dependence of 
the fields $\psi$ and $h$ on $\varphi$). Below we limited ourselves 
to the consideration of instabilities which do not lead to the 
breakdown of the azimuthal symmetry. For this reason we 
ignored this term.
\par
Eqs. (37) - (38) without a second term in the right-hand side (37) 
describe the dynamics of large-scale fluctuations in an incompressible 
conducting medium. If we make a linear analysis of  the  stability  of 
their solution,  we  can  reveal  that  it  is  unstable  and  we  can 
anticipate that the development of the  most  rapid  perturbations  is 
determined by a minimal number of modes, in particular by three.
\par
Let us show that the set of Eqs.(37) - (38) can be reduced  to  the  set 
of  three  nonlinear  differential   equations   that   describe   the 
interaction of the  three  perturbation  modes  (later  on  we  will 
discuss the plausibility  of this suggestion). Similar to 
[5], [10], we 
restrict ourselves to one perturbation mode for $\psi$ and to two modes for 
$h$ and also, following Lorenz [10], we assume free 
boundary  conditions 
for $\psi$: $\psi(0,z,t) = \psi(r_0,z,t) = 0$. Retaining the 
less significant terms 
in the Fourier representation of $\psi$ and $h$,  we  use  the  substitution, 
which transforms into the Lorenz substitution in case of  a  Cartesian 
coordinate system:
\begin{equation}
{\psi}(r,z,t) = \sqrt{2}{\frac{1+(\frac{{\pi}k}{g_1})^2}{k}}{{\nu}_m}
{X(t)}{\sin {(\frac{{\pi}kz}{r_0})}{J_1(\frac{{g_1}r}{r_0})}}
\end{equation}
\begin{equation}
h(r,z,t) =
\frac{H_0}{{\pi}r_1}\bigl(\sqrt{2}{Y(t)}{\cos {(\frac{{\pi}kz}{r_0})}}
{J_1(\frac{{g_1}r}{r_0})}-2Z(t){J_0(\frac{{g_1}r}{r_0})}
{J_1(\frac{{g_1}r}{r_0})}\bigr).
\end{equation}    
In (39) - (40) $g_1 = 3.83171$, corresponds to the first zero 
of the Bessel function 
${J_1}(x)$, $R_c = 64{g_1}^2{\pi}^2(b (4-b))^{-1}$ is the
critical Rayleigh number, $b = 4(1+({\pi}k{g_1}^{-1})^2)^{-1}$. 
By substituting Eqs.(39) -(40) for Eqs.(37) - (38), we find a set of 
ordinary differential equations for the $X(t), Y(t)$ and $Z(t)$ 
amplitudes, which coincides with the set of Eqs.(4) in 
[5] at $i(t) = I_0 = const$ 
(note that in [5]  $Z{J_1}(2{g_1}r{r_0}^{-1})$ is used for 
$h$ instead of 
$2Z{J_0}({g_1}r{r_0}^{-1}){J_1}({g_1}r{r_0}^{-1})$; 
this, however, does not transform  the  system 
of amplitudes obtained in [5], as substitution (21b), 
corresponding to 
the approximate equality ${J_1}(2x)\cong{2{J_0}(x){J_1}(x)}$, 
was  actually  used  in 
their derivation):
\begin{eqnarray}
{\dot X} & = & s(-X+IY),         \\
{\dot Y} & = & {\pi}{{g_1}^{-1}}X(-Z+{\pi}{{g_1}^{-1}}{r_1}I)-Y,   \\
{\dot Z} & = & -({\pi}{g_1}^{-1}XY+bZ).
\end{eqnarray}       
where the point symbol $"\cdot"$ is used to denote the  
differentiation  operator  for 
the dimensionless time $\tau =
4{g_1}^2b^{-1}{\nu_m}t{r_0}^{-2}$, $I(t) = i{I_0}^{-1}$  is the 
dimensionless current; $r_1 = R{R_c}^{-1}$ is the control 
parameter of the model. 
\par
The knowledge of $X(t), Y(t), Z(t)$ and $I(t)$ allows us to determine 
paths of particles transferring the mass (of "hydrodynamic" particles, 
below called atoms) and the current  (of  "conducting  electrons"),  while 
solving the motion equation: 
\begin{eqnarray}
{{\dot R}_a} & = & -{C_r}X{\cos ({\pi}k{Z_a})}{{J_1}({g_1}{R_a})}, \\
{{\dot Z}_a} & = & {C_z}X{\sin ({\pi}k{Z_a})}{{J_0}({g_1}{R_a})}
\end{eqnarray}       
and 
\begin{equation}
{{\dot R}_e} = {\sqrt{2}}Bk{r_1}^{-1}Y
{\sin ({\pi}k{Z_e})}{J_1}({g_1}{R_e}),
\end{equation}
\begin{eqnarray}
{{\dot Z}_e} & = & B(2I+a{r_1}^{-1}{\bigl(}{\sqrt{2}}Y
{\cos ({\pi}k{Z_e})}{{J_1}({g_1}{R_e})}+ \nonumber\\
& & +2Z({{J_1}^2({g_1}{R_e})}-{{J_0}^2({g_1}{R_e})}){\bigr)},
\end{eqnarray}                                                                 
where $R_a$, $Z_a$ and $R_e$, $Z_e$ are the coordinates 
of  paths  for  atoms  and 
conducting electrons respectively; $C_r =
2^{1/2}{\pi}{g_1}^{-2}$; $C_z = {g_1}k^{-1}{C_r}$; 
$B = bc{H_0}(16{\pi}e{n_e}{\nu_m}{g_1}^2)^{-1}$; 
$a = {g_1}{\pi}^{-1}$; $e$ and $n_e$ are, respectively, 
the unit charge and the density of conducting electrons. 
Since Eqs. (44) - (47) 
determine paths of particles corresponding to 
every moment of time $\tau$, 
then in their integrating the $X$, $Y$, $Z$ and $I$ amplitudes should be
assumed to be constant, i.e. systems (44) - (45) and 
(46) - (47) represent on the plane  $(r,z)$ 
autonomous dynamic systems of the second order. Moreover, the  set  of 
Eqs.(41) - (43) combined with additional equations 
for the  current  $I$ is  a 
control dynamic system for sets (44) - (45) and  (46) - (47).  
Thus  we  distinguish 
between "slow" processes, the  dynamics  of  which  is  determined  by 
system (41) - (43), and "rapid" ones, the dynamics of 
which is determined  by 
systems (44) - (45) and (46) - (47). Therefore, at certain 
values of $X$, $Y$, $Z$  and  $I$ 
in dynamic systems (44) - (45) and (46) - (47), 
one should  anticipate  bifurcations 
which can result in a topological rearrangement of spatial structures, 
in particular in a spontaneous breaking symmetry. 
\par
To find the value of current in the conductor, we should  specify 
the method of calculating the voltage drop across  it.  Experimentally 
the voltage drop is measured by determining the alteration in a 
magnetic flux coupled with the conductor {\({u_L} =
{L_{p0}}c^{-2}{\frac{di}{dt}}\)} ($L_{p0}$ is  the 
external inductance of the conductor) and also the Ohmic  voltage  drop 
that represents an integral of $z$-component of the  electric  field  on 
the conductor surface {\(E_z({r_0},z)\)}): {\({u_R} =
{\int_{0}^{l}{{E_z({r_0},z){dz}}}}\)}.  Then  with  the 
consideration of the boundary conditions, the voltage drop across  the 
conductor is 
\begin{equation}
u(t) = {u_L(t)}+{u_R(t)} = {L_{p0}}{I_0}({c^2}{t_0})^{-1}{\dot I}+
{R_{p0}}{I_0}(I-({\pi}{r_1})^{-1}{{J_0}^2(g_1)}Z).
\end{equation}
In Eq.(48) $t_0  = {{r_0}^2}b(4{g_1}^2{\nu_m})^{-1}$ 
is the base time; $R_{p0} = l({\pi}{r_0}^2{\sigma})^{-1}$ is 
the initial conductor resistance. It is evident that to  close  system 
(41) - (43), it is necessary to specify an equation (or equations) for 
determining the current $I$, the form of which 
depends on a concrete topology 
of an external circuit that serves as a thermostat with respect to the 
dynamic system (41) - (43). 
\section{Discussion}
\par
If we consider that ${\pi}{g_1}^{-1}\cong1$,  then  
Eqs.(41)  and  (42)  fully 
agree with the first two equations  in  the  Lorenz  model 
[10],  and 
Eq.(43) has the sign $"-"$ before $XY$ instead of the  
sign  $"+"$  in  the 
corresponding Lorenz equation. Thus our model describes NPT, as 
will be shown below.
\par
Let us find a spatial scale of perturbation in the  $z$  direction, 
which allows us to discuss the suitability of the assumption made when 
deriving Eqs.(41) - (43), that the minimal number of  
perturbation  modes  is 
three. We make use of the dependence of the critical  Rayleigh  number 
on the parameter $b$ (see also Fig.1): $R_c = 64{g_1}^2{\pi}^2
({b^2}(4-b))^{-1}$. We can 
see that at ${R_c}{\rightarrow}{\infty}$ at $b{\rightarrow}0$ or 
$b{\rightarrow}4$. The case $b = 0$ corresponds to a 
short-wave limit, i.e. $k = {\infty}$ ($\lambda = 0$). The case 
$b = 4$ corresponds to  a 
long-wave limit when $k = 0$ and $\lambda = \infty$. Hence, 
in the latter  case,  the 
perturbation spreads over the whole volume occupied by the matter.  In 
our case, a characteristic scale  of  the  problem  is  the  conductor 
radius $r_0$. Therefore it should  be  expected  that  the  most  rapidly 
developed perturbations are those with $\lambda \cong r_0$. 
\par
Actually, $R_c$ has a  minimum  corresponding  to  the  condition 
$dR_c/db = 0$: $b = 8/3$, $R_{c,min} = 978.08144$. This value  
of  the  Rayleigh 
critical number corresponds to the  perturbation  with  
$\lambda = {r_0}k^{-1} = {r_0}{\pi}{g_1}^{-1}b^{1/2}(4 -
b)^{-1/2} = 1.15931r_0$. Respectively, 
$r_1 = 1.0224\cdot10^{-3}{R_c} = 6.5088\cdot10^{-4}{I_0}^2
({c^2}{\rho_0}{\nu_m}\nu)^{-1}$. A subsequent 
perturbation evolution is possible in the direction of 
splitting the spatial 
scale. In addition, as shown in Fig. 1, the process of  splitting  the 
conductor into strata having the size $l = 2{\lambda}$  
(the  Rayleigh  critical 
number for a perturbation with the wave number $0.5k$ is 
$1.65\cdot10^3$ and 
for a perturbation with $l = 0.5\lambda$, it is  $1.957\cdot10^3$) 
is more advantageous at first. This analysis is supported 
by a  direct  observation 
of the conductor stratification in EEC  experiments  
[3], [4]  (Fig.2  in 
[11] shows that in the final explosion 
stage the conductor splits into 
strates with the size $l \cong 2\lambda$). In experiments 
with  copper  conductors 
0.58 mm in diameter [4], the mean distance 
between  the  strates  was 
0.78 mm. According to our model, $l \cong 2\lambda = 0.672$ mm, 
which  is  in  a 
rather good agreement with experiment (we did not use any experimental 
data while calculating $\lambda$). The consideration of  a  
larger  number  of 
perturbation modes does not show a significant change in  the  results 
of the above analysis, as modes with the wave number not equal 
to  $k$ 
have a longer time of perturbance than the  dominant  mode.  The  fact 
that $R_{c,min}$ corresponds to $\lambda \cong r_0$ 
shows  evidence  that  the  model 
proposed is a model of a non-equilibrium phase  transition  where  the 
amplitudes $X(t)$, $Y(t)$, $Z(t)$ and $I(t)$ have a 
physical  meaning  of  the 
parameters of order, the interaction of which is 
described by Eqs.(41) - (43) 
followed by one or several equations  for  finding  
the  current  $I(t)$ 
(depending on a concrete topology of an external electric
circuit). 
\par
Let us analyze the set of Eqs.(41) - (43) at $I = 1 = const$. We
have 
\begin{equation}
{\dot{\bss{W}}} = {\bss{F}}({\bss{W}}),\quad 
{\bss{W}} = \left( \begin{array}{ccc}
X \\
Y \\
Z 
\end{array} \right),
\end{equation} 
\begin{sloppypar}
The fixed points of the set  (49)  are  determined  by  the  condition 
${\bss{F}}({\bss{W}}) = {\bss{0}}$:  ${\bss{W}}_1 = {\bss{0}}$,  
${\bss{W}}_{2,3} = \{{\pm}1.221[b(1 - 0.672{r_1})]^{1/2}; 
{\pm}1.221[b(1 - 0.672{r_1})]^{1/2}; -1.221(1 - 0.672{r_1})\}$.
One can  see  that 
at $r_1 > r_*  = 1.488$, the set (49) has no fixed points. The analysis  of 
the solution stability shows that at $r_1 < r_*$ the  fixed  points  are 
stable. The fixed point corresponding to $r_1 = r_*$ is  a  stability 
boundary (Fig.3a shows a bifurcation curve for the set (49)). 
\end{sloppypar}
\par
The coordinates of the fixed  points  of  the  Lorenz  model  are 
${\bss{W}}_1 = {\bss{0}}$, ${\bss{W}}_{2,3} = \{{\pm}[b({r_1} - 1)]^{1/2}; 
{\pm}[b({r_1} - 1)]^{1/2}; {r_1} - 1\}$. The critical 
point of the Lorenz model ($r_1 = R{R_c}^{-1} = 1$)  
is  also  a  stability 
boundary (Fig. 3b shows a bifurcation curve of the Lorenz model).  One 
can see that at $R > R_c$  there are stable steady-state solutions in  the 
Lorenz model; a transition from the point  ${\bss{W}} = {\bss{0}}$ 
to  any  arbitrary 
point of the phase space being performed via the adiabatic sequence of 
stable states. A bifurcation, called  in  literature  a  supercritical 
bifurcation (the Hopf bifurcation), occurs at a critical point
[12]. 
\par
In our case, the bifurcation is a subcritical one, as at $R > R_c$  
there are no  steady-state  solutions  near  the  critical  point.  In 
addition, any trajectory of the system in a phase space at $R > R_c$  
shifts the system from the point ${\bss{W}} = {\bss{0}}$ to the 
infinity during a finite 
time. In literature this transition is called the explosive
[12]. 
\par
An asymptotical behavior of the solution of  set  (49)  is  of  a 
singular character:
\begin{equation}
X \sim {a_1}(t_* - t)^{-1},\quad 
Y \sim {a_2}(t_* - t)^{-2},\quad
Z \sim {a_3}(t_* - t)^{-2},
\end{equation} 
where $a_1 = {\pm}2.442$, $a_2 = {\pm}2.442s^{-1}$, $a_3 =
-2.442s^{-1}$. The Lorenz  model 
has a similar asymptotical behavior though with different from Eq.(50) 
factors: $\{a_i, i = 1,2,3\}$ [13]: 
$a_1 = 2i$, $a_2 = -2is$, $a_3 = -2s$ (an 
imaginary unit in factors a  and a indicates a  fluctuating  character 
in a steady-state solution of the Lorenz model). 
\begin{sloppypar}
It is worth while  coming  back  to  the  interpretation  of  the 
Rayleigh magnetic  number  and  the  control  parameter 
$r_1 = R{R_c}^{-1}$: $R = {v_A}^2{r_0}^2({\nu_m}\nu)^{-1}$. 
Sausage-type MHD  instabilities  are  known  to  be 
an alternative mechanism when a conductor is  destructed  by  current, 
the increment of which is ${\gamma}_A = {t_A}^{-1} =
2^{1/2}{v_A}{r_0}^{-1}$ ($t_A$ is the  time  for 
the development of a MHD instability) [14]. 
Then the Rayleigh 
magnetic number can be represented as a square ratio of 
the  times $t_d$ and 
$t_A$: $R = ({t_d}{t_A}^{-1})^2$ ($t_d =
{r_0}^2(2{\nu_m}\nu)^{-1}$ is the  effective  diffusion time). 
For $I = 1$, the equality $R = 1.488R_c$  corresponds 
to the critical  point 
on  the  bifurcation  curve,  where  as  the   
inequality $t_d{t_A}^{-1}\geq(1.488R_c)^{1/2} = 38.1495$ 
corresponds to the  supercritical  regime  in 
our model. 
\end{sloppypar}
\par
On the other hand sausage-type MHD  instabilities  are  known  to 
develop when the magnetic pressure $P_m  = H^2(8\pi)^{-1}$ 
surpasses the 
pressure in the medium. Let us estimate critical values  of  the  magnetic 
intensity at a constant current. The first value $H_{cr1}$ 
is determined as 
corresponding to the Rayleigh critical number 
$R_c = {H_{cr1}}^2(2{\pi}{\rho_0}{\nu_m}\nu)^{-1} = 
978.08144$: $H_{cr1} = 78.393{r_0}^{-1}({\nu_m}\nu)^{1/2}$. 
Setting $\sigma \cong 10^{17}\quad c^{-1}$,  we 
obtain $H_{cr1} = 2.1\cdot10^3{r_0}^{-1}{\eta}^{1/2}$ Oe. 
The second  value  of  the  critical 
field  intensity  ($H_{cr2}$)  is  determined  from  
the  pinch   condition 
${H_{cr2}}^2(8{\pi}P)^{-1}\geq1$. Setting
$P\cong{\rho_0}{c_s}^2$, $\rho_0\cong10\quad gcm^{-3}$, 
$c_s\cong10^5\quad cms^{-1}$, we 
obtain $H_{cr2} = 1.585\cdot10^6$ Oe. 
\par
Thus, at the conductor radius $r_0 = 10^{-2}$ cm widely used in 
experiment and the overrated shear viscosity factor 
${\eta}\cong1$ P, $H_{cr1} < H_{cr2}$. 
\par
The control parameter $r_1 = R{R_c}^{-1} = (H{H_{cr1}}^{-1})^2 >
1.488$. Hence,  in 
our model a non-equilibrium phase transition develops 
under the condition
$H\geq2.56165\cdot10^3{r_0}^{-1}{\eta}^{1/2}$ Oe. At 
$r_0\cong10^{-2}$ cm and $\eta \cong1$ P (six!) the 
following inequality is valid: 
$$2.56165\cdot10^5\quad{\rm Oe} < H < 1.585\cdot10^6\quad{\rm
Oe}$$.  
\par
Let us find the conductor radius ($r_*$), when 
$H_{cr1} = H_{cr2} = H$.  We 
obtain $r_* = 1.616\cdot10^{-3}{\eta}^{1/2} < 1.616\cdot10^{-3}$
cm. Since  the  actual  liquid 
metal viscosity is about $10^{-2}$ P, our  calculations  
are  overestimated 
at least by an order of magnitude. 
\par
Consequently, the instability  described  by  our  model  can  by 
refered to a class  of  magnetohydrodynamic  instabilities,  with  its 
excitation threshold being sufficiently lower than  that  of  ordinary 
sausage-type instabilities and therefore to investigate the former, it 
is not necessary, unlike in [14], to disturb  
the  conductor  surface. 
Moreover, vortex structures developed in the  conductor  during  their 
further evolution are acting on the conductor surface 
to  be  drawn  in,  which 
results in the conductor splitting (to describe the initial  phase  of 
splitting, we suggested a three-mode model  
[15],  similar  to  (41) - (43), 
according to which the $X$, $Y$ and $Z$ amplitudes 
have a time  singularity: 
$X \sim Y \sim Z \sim (t_* - t)^{-1/2}$ and $I \sim (t_* -
t)^{1/2}$; it should be noted  that 
the dynamics of the process in this stage is quantitatively similar to 
the destruction of superconductivity by a  critical  current).  As  an 
additional argument in  favor  of  the  transition  to  splitting  the 
following can be stated: at a constant current  the power $P = UI$  has 
the asymptote $P\sim(t_* - t)^{-2}$, in splitting 
it is $P\sim(t_* - t)^0 = const$.  Thus, 
it is more energetically beneficial to split a spatial scale  than  to 
increase the $X$, $Y$ and $Z$ amplitudes preserving 
its initial value of the scale. 
\section{Conclusions:}
\par
The principal results of the present paper can be summarized 
as follows: 
\par
- The initial model of the magnetic hydrodynamics of a  
conducting fluid has been transformed to the form where the  possibility  of 
large-scale structure formations is presented in an  explicit  
form, both the "heavy" and the "light" fluid (conducting electrons) consists 
of two fluids: laminar  and  vortex.  The  increase  in  an  effective 
conductor resistance even in the case of a constant local  conductance 
is a consequence of the vortex structure  formation  in  the  electron 
subsystem (electric current). 
\par
-  In  the  framework  of  the  present  model,  a  model  of   a 
non-equilibrium phase transition has been derived where the motion  of 
a laminar component of a heavy fluid was not taken  into  account  and 
kinetic coefficients were considered as constant, i.e. independent  of 
density and temperature. This model can be referred to a class of  the 
Lorenz model [10], the spatial scale of a vortex  
structure  being  in good agreement with experiment of [4]. 
\par
- The instability discussed in the paper has been shown to belong 
to a class of magnetohydrodynamic instabilities. It  is  characterized 
by a threshold current which is by a factor of $10^2$  
as a minimum lower than that for an ordinary sausage-type 
instability. There is also no 
need of a disturbance of the conductor surface to study this 
instability. Moreover, in case of a free conductor surface, the 
spatial scale splitting has been shown inevitable as a consequence of 
the initial vortex structure formation and evolution. 
\par
It should be noted that from the moment when  the  spatial  scale 
starts splitting, the model presented by Eqs.(41) - (43) loses 
its applicability. Therefore, our further problem is to develop 
models for a sequence of doubling the spatial scale, mentioned in section 3: 
{\({k_0}{\rightarrow}{k_1}=0.5{k_0}{\rightarrow}
{k_2}=2{k_0}{\rightarrow}{k_3}=2{k_2}{\rightarrow}
{k_4}=2{k_3}\)}, etc. In the stage  
${k_0}\rightarrow{k_1}$, as shown in [15], 
we can also be confined to three modes of perturbation.
\ack 
\par
We wish to thank Professors Sidorov A  F,  Zeldovich  B  Ya  and 
Fridman A M for their interest and stimulating discussion. One of the 
authors (Iskoldsky A M) is thankful to Professor  Zaslavsky  G  M for 
his advise to consider three-mode interaction. 
%\newpage 
\references
%\begin{thebibliography}{99} 
%\bibitem[1]{K-80} 
\numrefbk{[1]}{Klimontovich~Yu~L 1980}{Statistical Physics}{(Nauka: Moscow).}
%\bibitem[2]{H-88}
\numrefbk{[2]}{Haken~H 1988}{Information and Self-Organization. 
A Macroscopic Approach to Complex Systems}{(Springer Verlag: Heidelberg).}
%\bibitem[3]{L-85}
\numrefjl{[3]}{Lebedev~S~V~and~Savvatimskii~A~I 1985}{Sov. 
Phys. Usp.}{144}{243.}
%\bibitem[4]{I-85}
\numrefbk{[4]}{Iskoldsky~A~M 1985}{Thesis}{(High Current 
Electronics Institute: Tomsk) (in Russian).}
%\bibitem[5]{V-90}
\numrefjl{[5]}{Volkov~N~B~and~Iskoldsky A M 1990}{JETP Lett.}
{51}{634.}
%\bibitem[6]{dau-60}
\numrefbk{[6]}{Landau~L~D~and~Lifshitz~E~M 1960}
{Electrodynamics of Continuous Media}{(Pergamon Press: London).}
%\bibitem[7]{Le-37}
\numrefjl{[7]}{Leontovich~M~A~and~Mandelstam~L~I 1937}{JETP}
{7}{438 (in Russian).} 
%\bibitem[8]{I-83}
\numrefbk{[8]}{Iskoldsky~A~M~and~Romensky~Ye~I 1983}
{Dynamic Model of Thermoelastic Medium with Pressure 
Relaxations}{(Institute of Nuclear Physics: 
Novosibirsk) preprint N 83-11 (in Russian).}
%\bibitem[9]{S-62}
\numrefjl{[9]}{Saltzman~B 1962}{J. Atmos. Sci.}{19}{329.}
%\bibitem[10]{Lor-63}
\numrefjl{[11]}{Lorenz~E~N 1963}{J. Atmos. Sci.}{20}{130.}
%\bibitem[11]{B-79}
\numrefbk{[12]}{Baikov~A~P~et~al 1979}{Electrical Explosion of 
Conductors. 9. On Destortion of Conductors at 
the Electrical  Explosion}{(Institute 
of Automatic and Electrometry: Novosibirsk) 
preprint  N100  (in Russian).}
%\bibitem[12]{R-81}
\numrefbk{[12]}{Richtmyer~R~D 1981}{Principles of Advanced Mathematical 
Physics}{(Springer: New York) vol. 2.}
%\bibitem[13]{Lev-81}
\numrefbk{[14]}{Levine~G~and~Tabor~M 1981 in}{Progress in 
Chaotic Dynamic,}{eds. H. Flaschka and B. Chirikov 
(North-Holland: Amsterdam).}
%\bibitem[14]{A-75}
\numrefjl{[14]}{Abramova~K~B, Zlatin~N~A and 
Peregood~B~P 1975}{JETP}{69}{2007 (in Russian).}
%\bibitem[15]{V-91}
\numrefbk{[15]}{Volkov~N~B and Iskoldsky~A~M 1991 in}
{Proc.  8th  All-Union 
Conf. on Physics of Low-Temperature Plasma}{(Institute  of 
Physics Belorussia Academy of Science: Minsk) vol. 1 (in 
Russian).}
%\end{thebibliography}
\newpage
\figures
\figcaption{The magnetic Rayleigh number as a function 
of the parameter $b$.} 
\figcaption{X-ray patterns of an exploding  conductor  at  different  times 
(the experiment in [11]).} 
\figcaption{The bifurcation curves for our model ($a$)  and  for  the  Lorenz 
one ($b$).}

\end{document}